\documentclass[aps,prd,english,floatfix,reprint,twocolumn,showpacs,superscriptaddress,longbibliography,10pt,tightenlines]{revtex4-1}

\usepackage[utf8]{inputenc}
\usepackage{amssymb,amsmath}
\usepackage{graphicx}
\usepackage{natbib}
\usepackage{multirow}
\usepackage{braket}
\usepackage{subcaption}
\usepackage{ragged2e}
\DeclareCaptionJustification{justified}{\justifying}
\usepackage[final]{hyperref} 
\hypersetup{
	colorlinks=true,       
	linkcolor=red,        
	citecolor=red,        
	filecolor=magenta,     
	urlcolor=blue         
}
\usepackage{wasysym} 

\begin{document}
\title{Variations on vacuum decay: the scaling Ising and tricritical Ising field theories}
\author{M. Lencs{\'e}s}
\affiliation{Department of Theoretical Physics, Budapest University of Technology and Economics,\\ 1111 Budapest, M{\H u}egyetem rkp. 3, Hungary}
\affiliation{BME-MTA Statistical Field Theory ’Lend{\"u}let’ Research Group, Budapest University of Technology and Economics,\\ 1111 Budapest, M{\H u}egyetem rkp. 3, Hungary}
\author{G. Mussardo}
\affiliation{SISSA and INFN, Sezione di Trieste,\\ via Bonomea 265, I-34136, Trieste, Italy}
\author{G. Tak{\'a}cs}
\affiliation{Department of Theoretical Physics, Budapest University of Technology and Economics,\\ 1111 Budapest, M{\H u}egyetem rkp. 3, Hungary}
\affiliation{BME-MTA Statistical Field Theory ’Lend{\"u}let’ Research Group, Budapest University of Technology and Economics,\\ 1111 Budapest, M{\H u}egyetem rkp. 3, Hungary}
\affiliation{MTA-BME Quantum Dynamics and Correlations Research Group, Budapest University of Technology and Economics,\\ 1111 Budapest, M{\H u}egyetem rkp. 3, Hungary}

\date{2nd August, 2022}

\begin{abstract}
We study the decay of the false vacuum in the scaling Ising and tricritical Ising field theories using the Truncated Conformal Space Approach and compare the numerical results to theoretical predictions in the thin wall limit. In the Ising case, the results are consistent with previous studies on the quantum spin chain and the $\varphi^4$ quantum field theory; in particular we confirm that while the theoretical predictions get the dependence of the bubble nucleation rate on the latent heat right, they are off by a model dependent overall coefficient. The tricritical Ising model allows us on the other hand to examine more exotic vacuum degeneracy structures, such as three vacua or two asymmetric vacua,  which leads us to study several novel scenarios of false vacuum decay by lifting the vacuum degeneracy using different perturbations.
\end{abstract}
\maketitle

\section{Introduction}

The decay of the false vacuum is a fundamental and paradigmatic prediction of quantum field theory since the ground-breaking work by S. Coleman \cite{Coleman1977} which considered the process in a semi-classical approach, followed by including leading quantum corrections in \cite{Coleman1977-2}; for a very recent introduction on the topic see \cite{2022arXiv220503140D}. Such decay processes are hypothesised to have taken place in the early universe, and interest in this subject has been recently rekindled by the possible metastability of the electroweak vacuum \cite{Elias-Miro2012}.

Another reason for renewed interest on false vauum decay is provided by the advances of experimental techniques which promise to make accessible such a phenomenon in condensed matter laboratory experiments \cite{Billam2019,Billam2020,LunNg2020,Billam2021,Abel2021}: these advances have motivated recent theoretical studies both in the context of spin chains \cite{Rutkevich1999,2021PhRvB.104t1106L,Pomponio2022} and 1+1-dimensional $\varphi^4$ quantum field theory \cite{2022arXiv220515345S}, establishing that in one-dimensional systems the phenomenon can indeed be studied with sufficiently efficient methods to verify theoretical predictions for the bubble nucleation rate. These previous studies, however, were limited to the simplest scenario of an explicitly broken $\mathbb{Z}_2$ symmetry, where one starts from a model which has two degenerate vacua due to spontaneous symmetry breaking, with their degeneracy lifted by adding an explicit symmetry breaking external field.   

In this paper, the simplest scenario of false vacuum decay is realised in terms of the scaling Ising quantum field theory. However, we will go beyond that scheme by considering vacuum decay in relevant perturbations of the tricritical Ising conformal field theory, a model which allows us to realise  much more complex vacuum structures \cite{Lassig:1990xy} and therefore to induce various scenarios of vacuum decay. These include vacuum degeneracy unrelated to any spontaneous symmetry breaking, and also phases with three degenerate vacua. The scaling tricritical Ising field theory has recently been revisited by the authors of the present paper in relation, in particular, of two topics: 
the study of the Kramers-Vannier duality using the form factor bootstrap and the integrability of the model \cite{2022ScPP...12..162C}, and also the confinement phenomenon of the kink excitations into mesons \cite{2022PhLB..82837008L}. We note that kink confinement is another facet of lifting vacuum degeneracy, and so the study in this paper can also be viewed as complementary and a natural extension of \cite{2022PhLB..82837008L}. 

Following \cite{2021PhRvB.104t1106L,2022arXiv220515345S} we study the decay of the false vacuum as a quantum quench corresponding to a sudden change in the Hamiltonian of the quantum field theory. The resulting time evolution is simulated using the Truncated Conformal Space Approach (TCSA) \cite{Yurov_1989} from which we can  extract the dependence of the bubble nucleation rate on the latent heat, a quantity which we compare then with the theoretical predictions. 

The outline of the paper is as follows. Section \ref{sec:FalseVacuumDecay} describes the necessary theoretical background and the realisation of the false vacuum decay as a quantum quench. Section \ref{sec:isingdecay} addresses the simplest example of such a scenario, i.e. the case of the Ising field theory. We then turn to the description of various deformations of the tricritical Ising model and their vacuum structures in Section \ref{sec:vacua}. The numerical results obtained for various scenarios in the tricritical Ising model and their comparison to the theoretical expectations are presented in Section \ref{sec:TIM}, while Section \ref{sec:conclusions} contains our conclusions and outlook. Technical details of the TCSA simulations and the continuum limit of the bubble nucleation rate predicted for the spin chain are relegated to the Appendix.
 
\section{False vacuum decay: theory and quench protocol}\label{sec:FalseVacuumDecay}
\subsection{Theoretical predictions}
A metastable (false) vacuum state in quantum field theory decays via bubble nucleation initiated by quantum fluctuations: this is the scenario advocated by Coleman \cite{Coleman1977, Coleman1977-2}. In the semi-classical approximation, barrier penetration is dominated by the instanton bounce and the bubble nucleation rate, which is defined as the tunnelling rate per unit volume $V$, is given by
\begin{equation}\label{eq:coleman}
    \gamma=\frac{\Gamma}{V} = A \exp{\left[-\frac{1}{\hbar}S_{\text{E}}\right]}
\end{equation}
where $S_{\text{E}}$ is the Euclidean action of the instanton, while the prefactor $A$ can be computed as a determinant of quantum fluctuations in the instanton background: this latter quantity requires a careful treatment of zero modes which results in the tunnelling rate being proportional to the volume. 

Due to the energy cost of forming the walls (a.k.a.~surface tension), bubbles smaller than a critical size only appear as short-lived quantum fluctuations. Bubbles larger than the critical radius, however, undergo an accelerating expansion driven by the liberation of the latent heat $\Delta\mathcal{E}$, i.e., the difference between the energy densities of the false and the true vacua. 

In the thin wall limit of small $\Delta\mathcal{E}$, i.e., when the thickness of the walls is much smaller than the radius of the critical bubble, an explicit formula (which also includes the quantum corrections) was derived by Voloshin for 1+1 dimensional quantum field theories \cite{Voloshin1985}. In one spatial dimension, thin wall bubbles are eventually a kink-antikink pair with the true vacuum in their interior, and the critical diameter is
\begin{equation}\label{eq:resonant-bubble}
    a_* = \frac{2M}{\Delta\mathcal{E}}\,,
\end{equation}
where $M$ is the kink mass computed in the limiting case when the vacua are degenerate i.e. $\Delta\mathcal{E}=0$. Then the predicted nucleation rate is \cite{Voloshin1985} 
\begin{equation}\label{eq:voloshin}
    \gamma = \frac{\Delta\mathcal{E}}{2\pi} \exp{\left[-\frac{\pi M^2}{\Delta\mathcal{E}}\right]}\,.
\end{equation}
Similar results were obtained for tunnelling in the quantum Ising spin chain \cite{Rutkevich1999}. However, the continuum limit of the latter result (c.f. Appendix \ref{sec:contlimit})
\begin{equation}
    \gamma=\frac{\pi\Delta\mathcal{E}}{18}\exp\left\{ -\frac{\pi M^{2}}{\Delta\mathcal{E}}\right\}
\end{equation} 
differs from \eqref{eq:voloshin} by a dimensionless numerical coefficient, signalling somehow that our theoretical understanding of the nucleation rate is still incomplete. 

Indeed, the recent numerical simulation studies performed on the Ising spin chain \cite{2021PhRvB.104t1106L} and the $1+1$-dimensional $\varphi^4$ scalar field theory \cite{2022arXiv220515345S} confirm that the bubble nucleation rate is described by a formula whose  general expression is
\begin{equation}\label{eq:Cvoloshin}
    \gamma = \mathcal{C}\, \Delta\mathcal{E} \exp{\left[-\frac{\pi M^2}{\Delta\mathcal{E}}\right]}
\end{equation}
with a dimensionless coefficient $\mathcal{C}$ to be determined. Finally we note that in the $1+1$-dimensional $\varphi^4$ scalar field theory $\mathcal{C}$ turns out to depend on the self-interaction strength \cite{2022arXiv220515345S}.

\subsection{Vacuum decay as a quantum quench}\label{subsec:decayQQ}

Let's now turn to the description of the protocol used to investigate the vacuum decay, which is a quantum quench \cite{2006PhRvL..96m6801C} corresponding to a sudden change in the Hamiltonian at initial time $t=0$. The idea is to consider a system with degenerate vacua described by some action $\mathcal{A}_0$ and perturb it by means of an additional operator which lifts the degeneracy
\begin{equation}
\mathcal{A}_\varepsilon=\mathcal{A}_0-\varepsilon\int d^2x \, \Phi(x)\,. 
\end{equation}
For the corresponding Hamiltonian we have 
\begin{equation}
H_\varepsilon=H_0+\varepsilon\int dx \, \Phi(x)\,.
\end{equation}
When the vacuum degeneracy is a consequence of spontaneous symmetry breaking, $\Phi(x)$ can be chosen as a field which explicitly breaks the symmetry but, in the following,  we will also consider situations which are more general. Since we will work in the thin wall regime, $\varepsilon$ has to be considered small and therefore the latent heat can be computed in first order perturbation theory as
\begin{equation}
\Delta\mathcal{E}=\varepsilon \left(\Braket{+|\Phi(x)|+}-\Braket{-|\Phi(x)|-}\right)    
\end{equation}
where $\ket{+}$ and $\ket{-}$ denote respectively the false and true vacuum states in the limit $\varepsilon\rightarrow 0$. An important common aspect of the unperturbed models analysed in this paper (listed in Section \ref{sec:vacua}) is that all of them are integrable, and therefore the  expectation values of relevant perturbing fields are known exactly \cite{1997NuPhB.493..571L,1998NuPhB.516..652F}. In addition, due to their integrability, the various kink masses present in the different phases of the model reached in the $\epsilon=0$ limit are also exactly known  \cite{1994PhLB..324...45F}, which permits the explicit evaluation of the predicted bubble nucleation rate \eqref{eq:voloshin}.

The protocol is set up as follows. For a given value of $\varepsilon$, the initial state $\ket{\Psi(0)}$ is determined as the ground state for $-\varepsilon$, and is then evolved by the Hamiltonian for $+\varepsilon$:
\begin{equation}
\ket{\Psi(t)}=e^{-iH_\varepsilon t} \ket{\Psi(0)}\,.    
\end{equation}
To track the vacuum decay, it is necessary to compute the time evolution of some observable:
\begin{eqnarray}
  \mathcal{O}(t)=\braket{\Psi(t)|\mathcal{O}|\Psi(t)}
\end{eqnarray}
For instance, in the simple case of vacua corresponding to a broken $\mathbb{Z}_2$ symmetry, the observable can be chosen as the corresponding order parameter $\sigma$, since by definition its sign distinguishes between the vacua:
\begin{equation}
  \Braket{+|\sigma|+}=-\Braket{-|\sigma|-}\quad\text{for}\quad\varepsilon=0\,.
\end{equation}
For any suitable observable ${\mathcal O}(t)$, the combination
\begin{equation}
    f^{\mathcal{O}}(t) = \frac{\langle\mathcal{O}(t)\rangle+\langle\mathcal{O}(0)\rangle}{2\langle\mathcal{O}(0)\rangle}
    \label{eq:fO}
\end{equation}
is a very convenient quantity, since initially it satisfies $f^{\mathcal{O}}(0)=1$ and (neglecting corrections due to explicit symmetry breaking $\varepsilon\neq 0$) it vanishes in the true vacuum, so the transition corresponds to $f^{\mathcal{O}}(t)$ changing from $1$ to $0$. 
For the sake of uniformity we track the evolution of the combination \eqref{eq:fO} in all cases, including those where the vacuum structure is not determined by a broken $\mathbb{Z}_2$ symmetry.

The time evolution of the system consists of three regimes \cite{2021PhRvB.104t1106L,2022arXiv220515345S} which lead to the following behaviors of $ f^{\mathcal{O}}(t)$
\begin{itemize}
    \item Initial transient: for short times the evolution of $ f^{\mathcal{O}}(t)$ is quadratic
    \begin{equation}
        f^{\mathcal{O}}(t)-1\propto t^2+\dots
    \end{equation}
    corresponding to the quantum Zeno regime \cite{Degasperis1974, Misra1977}. 
    \item Nucleation: for intermediate times the evolution of $ f^{\mathcal{O}}(t)$ is dominated by the bubble nucleation rate and the time dependence has the form 
     \begin{equation}
        f^{\mathcal{O}}(t)\propto e^{-\Gamma t}\,,
    \label{eq:expregime}\end{equation}
    with $\Gamma=\gamma R$ in terms of the volume $R$ of the system.
     \item Thermalisation: for longer times the evolution of $ f^{\mathcal{O}}(t)$ becomes very complicated due to several processes including the expansion of nucleated bubbles and their collisions, ultimately leading to thermalisation of the system.
\end{itemize}

The above considerations imply that the decay rate of the false vacuum can be extracted identifying the  intermediate time regime where the exponential dependence \eqref{eq:expregime} holds. This can be found by analysing the time dependence of $f^{\mathcal{O}}(t)$ which we compute by simulating the time evolution using the Truncated Conformal Space Approach invented by Yurov and Zamolodchikov \cite{Yurov_1989}, and later applied to the scaling Ising \cite{1991IJMPA...6.4557Y}  and tricritical Ising \cite{Lassig:1990xy}. The numerical computations were carried out using a recently developed package which utilises the chiral structure of conformal field theory with periodic boundary conditions \cite{2022CoPhC.27708376H}.

The TCSA simulates the quantum field theory in finite volume and with an energy cut-off, which impose limitations on the range of latent heat $\Delta\mathcal{E}$ for which vacuum decay can be studied. In particular, the finite volume introduces a lower limit on $\Delta\mathcal{E}$ so that bubble nucleation is not affected by finite size effects, while the energy cut-off results in an upper limit on this quantity  \cite{2022arXiv220515345S}. A detailed discussion of these conditions is given in Appendix \ref{subsec:TCSA_conditions}. In addition, obtaining sufficiently precise results requires extrapolation in the cut-off. Presently, for non-equilibrium time evolution this can only be carried out partially, and our procedure together with its limitations are described in Appendix  \ref{subsec:approximation}.

\section{Warming up: vacuum decay in the Ising QFT}\label{sec:isingdecay}

\begin{figure}
	\centering
	\begin{subfigure}{0.5\textwidth}
     \centering
		\includegraphics{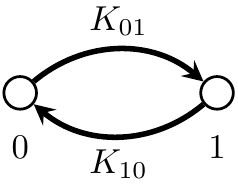}
		\caption{\label{fig:Isingkink}The two vacua in the ferromagnetic phase of the Ising model labelled by $0$ and $1$, with the kinks interpolating between them depicted as arrows.}
	\end{subfigure}
	\begin{subfigure}{0.5\textwidth}
	\centering
		\includegraphics{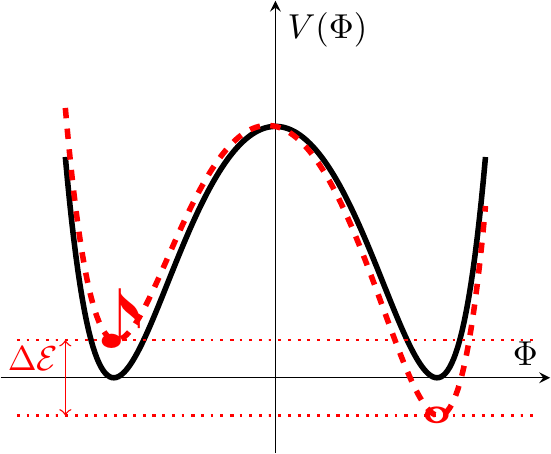}
    \caption{\label{fig:Ising_GL}Ginzburg-Landau potential for $h=0$ (solid black curve) and $h\neq 0$ (dashed red curve). The false/true vacua are indicated with the musical notes {\large \eighthnote}/{\large \fullnote} respectively, and the difference in their energy density $\Delta\mathcal{E}$ is also shown.}
	\end{subfigure}
\caption{(a) Vacua and kink structures, and (b) sketch of the Ginzburg--Landau potential in the thermal perturbation of the Ising, with $\Phi$ denoting the order parameter.}
\end{figure}

The simplest example of false vacuum decay is obtained considering the scaling Ising CFT of central charge $c=1/2$ with a Hamiltonian $H_{*}$, perturbed by the energy density operator $\epsilon$ with conformal weights $(1/2,1/2)$ leading to the Hamiltonian
\begin{equation}
    H_{0}=H_{*}-\frac{M}{2\pi}\int dx \, \epsilon(x)\,.
\end{equation}
The (-) sign in front of the perturbation corresponds to the ferromagnetic phase where the model has two degenerate ground states $\ket{\pm}$, which are connected by kinks/antikinks of mass $M$, as shown in Fig. \ref{fig:Isingkink}. The order parameter is the spin field $\sigma$ with conformal weights $(1/16,1/16)$, and its exact expectation value is 
\begin{equation}
    \braket{\pm|\sigma|\pm}=\pm\langle\sigma\rangle\quad , \quad\langle\sigma\rangle=2^{1/12}e^{-1/8}A^{3/2}M^{1/8}\,,
\label{eq:Isingsigmavev}
\end{equation}
where $A=1.2824271291\cdots$ is Glaisher's constant. Adding a nonzero magnetic field
\begin{equation}
    H_{h}=H_{0}+h\int dx \, \sigma(x)\,,
\end{equation}
lifts the degeneracy, leading to a system with a false and a true vacuum with the energy density difference:
\begin{equation}
    \label{eq:IsingEps}
    \Delta \mathcal{E} = 2h\langle\sigma\rangle = 2.7156766834\cdots h M^{1/8}\,,
\end{equation}
as shown in Figure \ref{fig:Ising_GL}. This model can be viewed as the continuum limit of the spin chain considered in \cite{2021PhRvB.104t1106L}, and it is in the same universality of the $\varphi^4$ theory for which the false vacuum decay was studied in \cite{2022arXiv220515345S}.
\begin{figure}[t]
    \centering
    \includegraphics{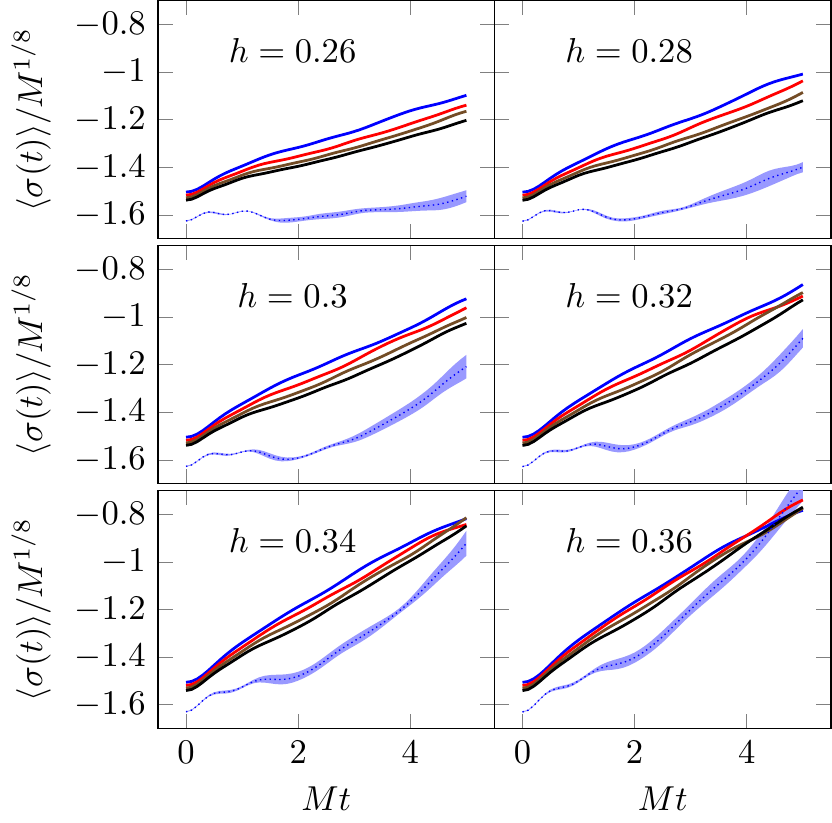}
    \caption{The time evolution of $\langle \sigma(t) \rangle$ in the Ising model for various $h\rightarrow -h$ quenches, ranging from $hM^{-15/8}=0.26$ to $0.36$ in dimensionless volume $MR=30$. Continuous lines are the raw TCSA data (blue, red, brown and black for cutoffs $16,18,20$ and $22$ respectively), while the dotted blue lines are extrapolated, with shaded areas indicating the uncertainty of the extrapolation.}
    \label{fig:SigmaVEV}
\end{figure}

\begin{figure}[t]
    \centering
    \includegraphics{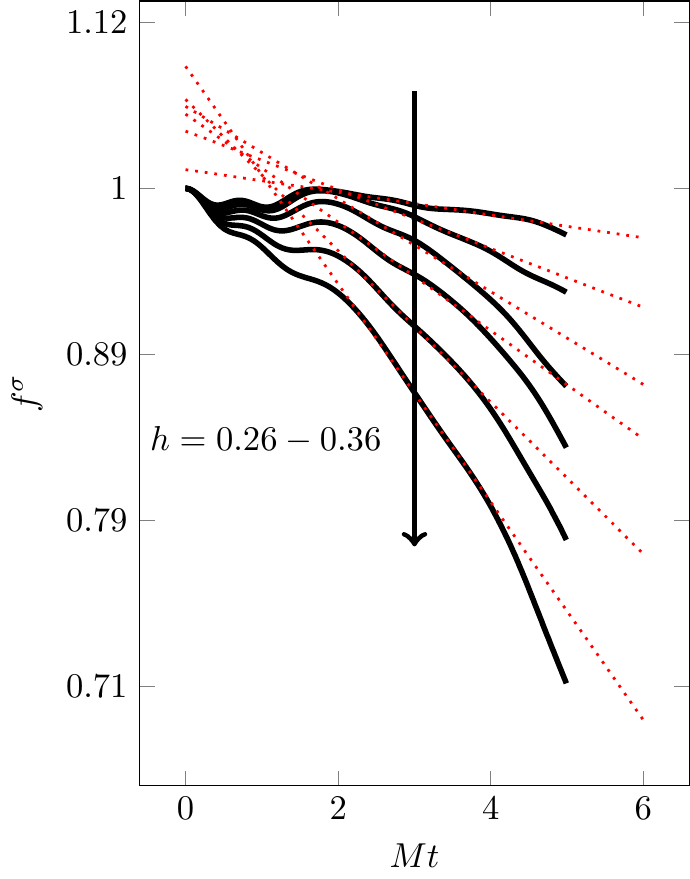}
    \caption{A few samples of the time evolution of the quantity $f^{\sigma}$ in the Ising model together with curves fitted to the exponentially decaying part for $MR= 30$ and for longitudinal fields $hM^{-15/8}=0.26,0.28,0.30,0.32,0.34,0.36$.}
    \label{fig:fsigmaIsing}
\end{figure}

\begin{figure}[t]
    \centering
    \includegraphics{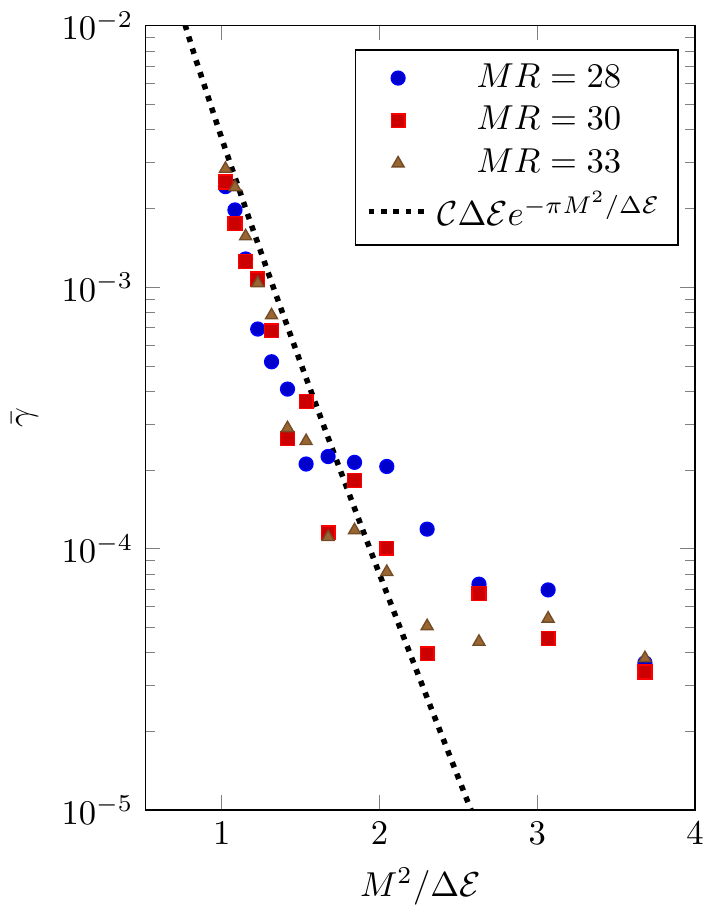}
    \caption{The dimensionless bubble nucleation rate $\bar\gamma$ in the Ising model extracted from the time evolution of the expectation value of $\sigma$, as a function of the latent heat in different volumes.}
    \label{fig:IsingSigmaGamma}
\end{figure}

As described in Subsection \ref{subsec:decayQQ}, we start the time evolution from  the ground state of the Hamiltonian with magnetic field $-h$ (this ground state determined numerically by TCSA) and then evolve the system by the Hamiltonian with field $+h$. The observable used to follow the time evolution is the order parameter $\sigma$, and the results are illustrated in Fig.~\ref{fig:SigmaVEV}. We extrapolated the time-dependent expectation value in the cut-off using the procedure introduced in \cite{Rakovszky_2016}; for a description and limitations of this approach c.f. Appendix \ref{subsec:approximation}. Using these data, the quantity 
\begin{equation}
    f^{\sigma}(t) = \frac{\langle\sigma(t)\rangle+\langle\sigma(0)\rangle}{2\langle\sigma(0)\rangle}
    \label{eq:fsigma}
\end{equation}
can be plotted on a logarithmic scale as shown in Fig.~\ref{fig:fsigmaIsing}, and the decay constant $\Gamma=\gamma \,R$ can be extracted from the flat region of the curves, as a function of $h$. 
Figure \ref{fig:IsingSigmaGamma} shows the resulting values for $\gamma$ as a function of $1/\Delta\mathcal{E}$ on a logarithmic scale. In dimensionless units, \eqref{eq:Cvoloshin} predicts the relation
\begin{equation}
    \log \bar{\gamma}=\log \mathcal{C}+\log\frac{\Delta\mathcal{E}}{M^2}-\frac{\pi M^2}{\Delta\mathcal{E}}
\label{eq:logpredictionC}\end{equation}
where 
\begin{equation}
    \bar{\gamma}=\frac{\gamma}{M^2}
\end{equation}
is the dimensionless bubble nucleation rate (in units of the sqaure of the kink mass $M$), and the prefactor $\mathcal{C}$ is $1/2\pi$ according to \eqref{eq:voloshin}. It turns out that the agreement is very good over more than one decade of $\gamma$ except for the value of the prefactor, which we used as a fitting parameter separately for data from each volume. The values obtained from the fit are $\mathcal{C}=0.053(5)$, $0.057(8)$ and $0.057(7)$ for the volumes $MR=28$, $30$ and $33$, showing no significant dependence on the volume. Note that the necessity of adjusting the prefactor as a fit parameter was already noticed in the case of the spin chain \cite{2021PhRvB.104t1106L} as well as for the $\varphi^4$ theory \cite{2022arXiv220515345S}, and that the volume independence is consistent with the results of the latter study. 

Deviations between the theoretical prediction (amended by fitting the prefactor) and numerical results for the nucleation rate are expected from finite size effects for small values of $\Delta\mathcal{E}$ such that the dimensionless size of the resonant bubble 
\begin{equation}\label{eq:resonant-bubble-dimless}
    Ma_* = \frac{2M^2}{\Delta\mathcal{E}}\,,
\end{equation}
is of order $MR$. However, the numerically observed deviations clearly visible in Fig. \ref{fig:IsingSigmaGamma} appear when the bubble size is still much smaller than the volume, and originate from the unreliability of extracting the slope from the time evolution which shows no discernible sign of exponential decay for too small values of $h$, c.f. Fig.~\ref{fig:fsigmaIsing}.

In the present case of the Ising model the quality of the data even allows us to  numerically determine the coefficient of the leading term of the $\Delta\mathcal{E}$ dependence in \eqref{eq:logpredictionC}. Fitting the data with the function
\begin{equation}
    \log \bar{\gamma}=c+\log\frac{\Delta\mathcal{E}}{M^2}-a \frac{M^2}{\Delta\mathcal{E}}
\label{eq:logfit}
\end{equation}
in terms of the parameters $c$ and $a$, the expected value $\pi$ for $a$ is reproduced reasonably well as shown in Table \ref{tab:pi}.

\begin{table}
\begin{center}
    \begin{tabular}{|c|c|c|c|}
         \hline
         $MR$ & $28$ & $30$ & $33$  \\
         \hline
         \hline
         $a$ &  $3.2(4)$ & $2.9(5)$ & $3.1(4)$ \\ 
         \hline
    \end{tabular}
             \caption{\label{tab:pi}Fitted values for the parameter $a$ in \eqref{eq:logfit}, for different values of the volume parameter $MR$. The expected result is $\pi$. }
\end{center}
\end{table}

\section{Vacuum structures in the tricritical Ising model}\label{sec:vacua}
Here we outline the vacuum structure of three different perturbations of the tricritical Ising model. For a general understanding of how these arise in perturbed CFT we refer the reader to \cite{2009JPhA...42D4022M}. We note that all these perturbed CFT are integrable, which allows us to extract the parameters needed as inputs for the theoretical prediction of the bubble nucleation rate.

\subsection{Thermal deformation}
The thermal perturbation is defined by the formal action
\begin{eqnarray}
  \mathcal{A}_{\epsilon}=\mathcal{A}_{TIM} + g \, \int d^2x\, \epsilon(x) \,, 
  \label{ttttherm}
\end{eqnarray} 
where $\mathcal{A}_{TIM}$ corresponds to the unique unitary conformal field theory with central charge $c=7/10$, and the perturbing field $\epsilon$ is the primary field of conformal weights $(1/10,1/10)$. This model is integrable and its spectrum consists of $7$ excitations, whose scattering is described by the $E_7$ S matrix \cite{MC,FZ}. 
This direction describes the phase transition related to the  $\mathbb{Z}_2$ symmetry, with $g$ positive / negative corresponding to the paramagnetic / ferromagnetic phase, respectively. In the ferromagnetic phase, $\mathbb{Z}_2$ is spontaneously broken, and the vacua are connected by topological excitations with the structure shown in Fig.~\ref{fig:E7kink}. The kink mass can be computed exactly in terms of the coupling constant~\cite{1994PhLB..324...45F}:
\begin{equation}
  M=3.745372836\dots\cdot|g|^{5/9}  \,.
  \label{eq:e7gap}
\end{equation} 
In the paramagnetic phase there is a single vacuum and all $7$ excitations are topologically trivial. We recently examined this model in detail \cite{2022ScPP...12..162C}, and the interested reader is referred to this work for more details and further references.

In the ferromagnetic phase the vacuum degeneracy can be lifted by adding either the leading magnetisation $\sigma$ with dimensions $(3/80,3/80)$, or the subleading $\sigma'$ with dimensions $(7/16,7/16)$, both of which lead to confinement of the topological excitations with similar phenomenology \cite{2022PhLB..82837008L}. 

\subsection{Sub-leading magnetization deformation}\label{subsec:sigmapvacs}

\begin{figure}
	\centering
	\begin{subfigure}{0.5\textwidth}
     \centering
		\includegraphics{E7.pdf}
		\caption{\label{fig:E7kink}Kink structure for the thermal perturbation, with the two vacua labelled by $0$ and $1$.}
	\end{subfigure}
	\begin{subfigure}{0.5\textwidth}
	\centering
		\includegraphics{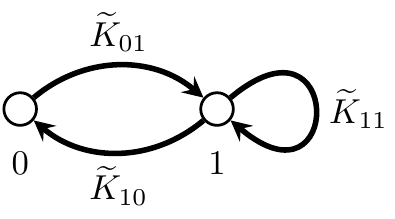}
		\caption{\label{fig:A3kink}Kink structure for the subleading magnetisation perturbation, with the two vacua labelled by $0$ and $1$.}
	\end{subfigure}
	\begin{subfigure}{0.5\textwidth}
	\centering
		\includegraphics{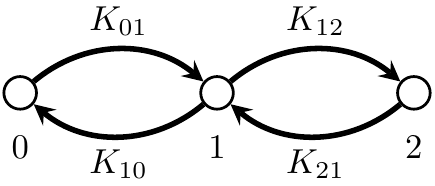}
		\caption{\label{fig:A4kink}Kink structure for the vacancy density perturbation, with the two vacua labelled by $0$, $1$ and $2$.}
	\end{subfigure}
\caption{Vacua and kink structures for different perturbations of the tricritical Ising model}
\end{figure}

Interestingly enough, perturbing the model by the subleading magnetisation operator of conformal weight $(7/16,7/16)$ as
\begin{eqnarray}
  \mathcal{A}_{\sigma'}=\mathcal{A}_{TIM} + h' \, \int d^2x\, \sigma'(x) \,, 
  \label{eqn:subleadingmagn}
\end{eqnarray} 
results in a phase of the model where there are two degenerate vacua despite the absence of any broken global symmetry. These two vacua are physically different, which is manifested also by the structure of kinks as shown in Fig.~\ref{fig:A3kink}, with their exact scattering amplitudes derived in \cite{1992IJMPA...7.5281C}, and the  exact relation of the kink mass to the coupling is given by~\cite{1994PhLB..324...45F}
\begin{equation}
    M=4.927791224\dots\cdot|h'|^{8/9}\,.
\label{eq:a3gap}
\end{equation}
We note that in this case the physical behaviour of the model is independent of the sign of the coupling $h'$.

The degeneracy between the vacua can be lifted by adding the thermal perturbation $\epsilon$, with very different effects depending on the sign of the thermal coupling, as described in relation to confinement in \cite{2022PhLB..82837008L}. The effect of the $\epsilon$ perturbation in the context of vacuum tunneling is discussed in Subsection \ref{subsec:sigmaprime}.

\subsection{Vacancy density deformation}

The perturbation by the vacancy density operator $t$ of conformal dimensions $(3/5,3/5)$
\begin{eqnarray}
  \mathcal{A}_{t}=\mathcal{A}_{TIM} + \mu \, \int d^2x\, t(x) \,, 
  \label{vacancypert}
\end{eqnarray} 
has a very different behaviour depending on the sign of the coupling:
\begin{itemize}
    \item For $\mu<0$ the system develops a mass gap, with the fundamental excitation being kinks interpolating between three degenerate vacua. The degeneracy is partially related to a broken $\mathbb{Z}_2$ symmetry which makes two of the vacua equivalent, while the third one is physically different. The kink structure is shown in Fig.~\ref{fig:A4kink} and the corresponding exact S-matrix can be obtained from a restriction of the sine-Gordon model \cite{Reshetikhin:1989qg,Bernard:1990cw}. The mass of the kinks interpolating between neighbouring vacua is related to $\mu$ as\cite{1995IJMPA..10.1125Z}
    \begin{equation}
    M = 10.829980\cdots |\mu|^{5/4}
    \label{eq:A4kinkmass}
    \end{equation}
    \item For $\mu>0$ the model describes a famous massless flow ending in the critical Ising CFT with central charge $1/2$ \cite{Zammassless}. The low energy excitations are massless kinks and the vacuum structure is similar to the one of the $E_7$ model shown in Fig.~\ref{fig:E7kink}.
\end{itemize}

\section{Vacuum decay in perturbations of the tricritical Ising CFT}\label{sec:TIM}

\subsection{Vacuum decay in the thermal deformation induced by magnetisation or subleading magnetisation}\label{subsec:E7}

\begin{figure}
    \centering
    \includegraphics{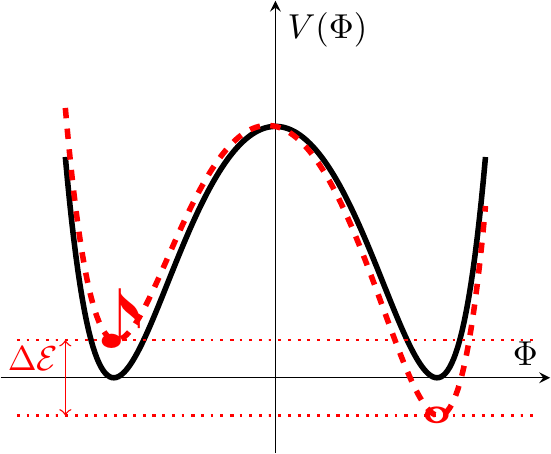}
    \caption{Sketch of the Ginzburg--Landau potential in the thermally perturbed tricritical Ising model (solid black curve) and its deformation induced by the magnetization (dashed red curve). The location of the false/true vacuum are denoted by {\large \eighthnote}/{\large \fullnote} respectively. The energy difference is also indicated between them.}
    \label{fig:E7sigmaGL}
\end{figure}

\begin{figure}
    \centering
    \includegraphics{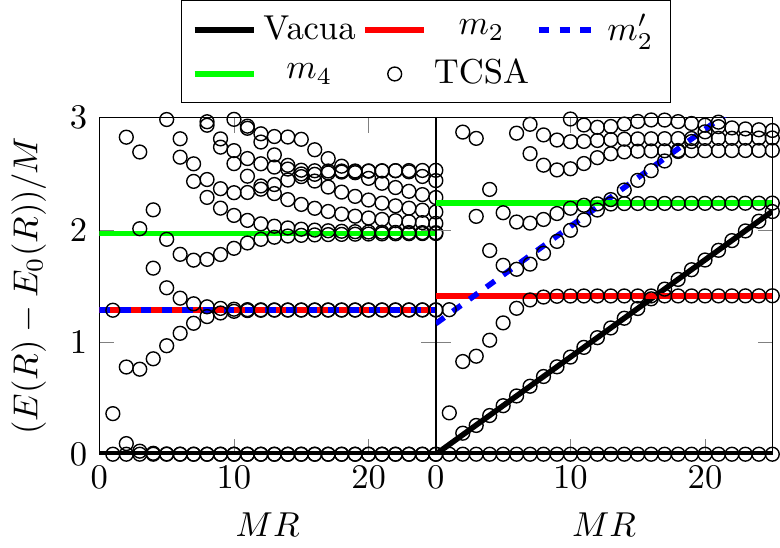}
    \caption{Finite volume energy levels in $H_{\epsilon}$ and its perturbation with $\sigma$, relative to the lowest level. \textbf{Left}: for $h=0$ there are two vacua which become degenerate for $R\rightarrow\infty$, and accordingly all even particles appear in two copies, as can be seen for the particles of mass $m_2$ and $m_4$. The red/green lines show the exact values of the masses predicted by the integrable $E_7$ scattering theory. \textbf{Right}: for $h>0$ the vacuum degeneracy is lifted and the false vacuum acquires a slope given by~\eqref{eq:falseslope} relative to the true vacuum, together with the particle excitations associated to them. The mass relative of the metastable particle excitation to the false vacuum was extracted from the frequency of the post-quench oscillations and is shown by the blue dotted lines. Note that this value fits very well the energy level corresponding to the metastable particle state, confirming the origin of the post-quench oscillations discussed in the main text. In addition, the masses of the stable particles also acquire corrections compared to their original values, as shown by the displacement of the red and green lines compared to the left panel. Particles on the true vacuum get mass corrections (fit to TCSA data indicated).}
    \label{fig:E7TCSAspectrum}
\end{figure}

The vacuum degeneracy in the model~\eqref{ttttherm} can be lifted by adding either the leading magnetisation $\sigma$ with dimensions $(3/80,3/80)$, or the subleading $\sigma'$ with dimensions $(7/16,7/16)$:
\begin{eqnarray}
  \nonumber H&=&H_\epsilon + h \int dx \,\sigma(x),\\
  H'&=&H_\epsilon + h' \int dx \,\sigma'(x).
\end{eqnarray}
For small couplings $h,h'$ (which can be assumed positive without loss of generality) both perturbations lead to the Ginzburg--Landau potential depicted for $h<0$ in Fig.~\ref{fig:E7sigmaGL}, which is qualitatively identical to the Ising case discussed in Section \ref{sec:isingdecay}. The latent heat is given by
\begin{eqnarray}
  \nonumber \Delta \mathcal{E} &=& 2h\langle \sigma \rangle,\\
  \Delta \mathcal{E}' &=& 2h'\langle \sigma' \rangle,
  \label{eq:falseslope}
\end{eqnarray}
with
\begin{align}
\braket{\sigma}&=1.5927\dots \cdot(-g)^{1/24} \nonumber\\
\braket{\sigma'}&=2.45205\dots \cdot(-g)^{35/72}\,,
\end{align} 
while the relation of the kink mass $M$ to the coupling $g$ can be found in Eq. \eqref{eq:e7gap}.
\begin{figure}
    \centering
    \includegraphics{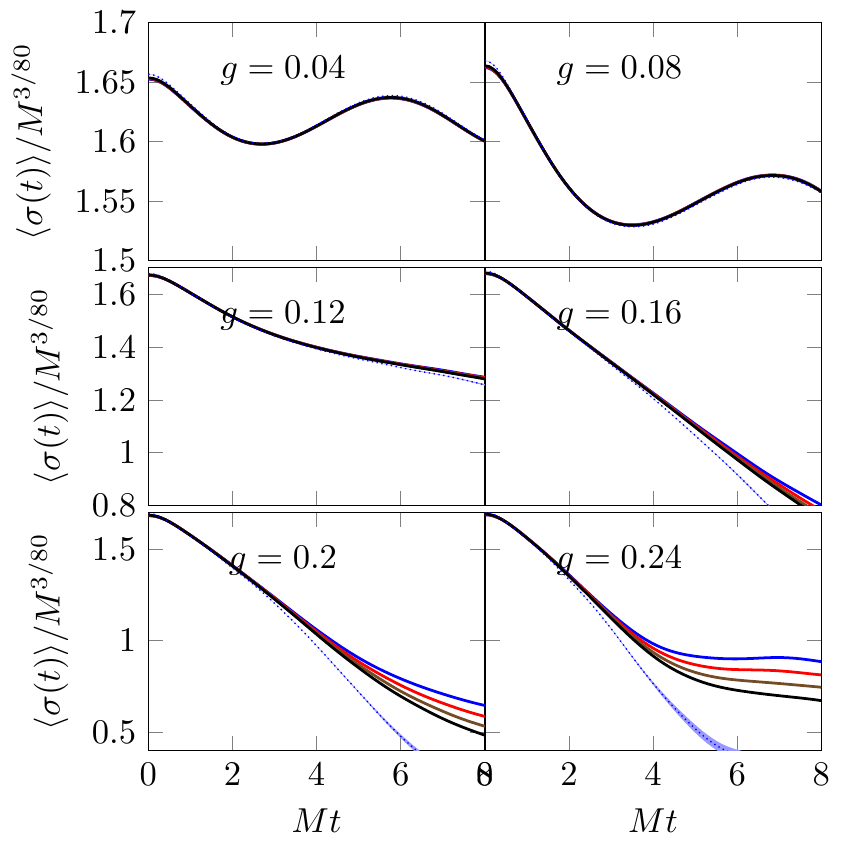}
    \caption{Extrapolation of the time evolution of the expectation value $\langle \sigma(t) \rangle$ calculated for quenches with flipping the sign of the magnetic field at $MR=35$. The extrapolation is very accurate in this case. However, it is impossible to reliably fit the exponential decay due to large oscillations.}
    \label{fig:E7timevol}
\end{figure}
The change in the (finite volume) spectrum induced by adding the perturbation $\sigma$ is shown in Fig. \ref{fig:E7TCSAspectrum}. Note the presence of degenerate vacua for $h=0$, all of which have a corresponding copy of the $\mathbb{Z}_2$ even particle excitations of mass $m_2=2M\cos 5\pi/18$ and $m_4=2M\cos \pi/18$ (there are no one-kink levels as these are excluded by periodic boundary conditions). After adding the perturbation, the false vacuum gains an energy which increases linearly with the volume signaling a metastable state, with the slope being the latent heat $\Delta\mathcal{E}$. Note that of the two copies of the one-particle excitations, one remains stable over the true vacuum, while the other gains a linear contribution identical to that of the false vacuum and becomes a metastable excitation.

Examples of the time evolution for the case of perturbation with the leading magnetisation $\sigma$ operator for various values of $h$ are presented in Fig.~\ref{fig:E7timevol}, where the quantity followed in time is the expectation value of the leading magnetisation operator $\sigma$. The numerical data obtained for different cut-offs can be reliably extrapolated using the procedure described in Appendix \ref{subsec:approximation}. For small values of $h$, one observes large oscillations, with the dominant frequency matching the value $m_2$, i.e. the mass of the lowest even particle. The origin of these oscillations is that the quench excites the metastable particle states over the false vacuum discussed above. The presence of such oscillations in expectation values of local operators can be regarded as a generic feature of quantum quenches when there is a one-particle contribution to the time evolution and it can be established by using first order perturbation theory \cite{2014JPhA...47N2001D,2017JPhA...50h4004D,2022NuPhB.97415643D}; such oscillations were also observed in the time evolution of entanglement \cite{2020PhRvL.124w0601C}. We note that, contrary to the results obtained in first order perturbation theory, these oscillations are in general exponentially damped, as shown by explicit simulations of the time evolution \cite{Rakovszky_2016}; however, the exponential damping can only be obtained by summing up the contribution of kinematic poles to all orders \cite{2012JSMTE..04..017S,2014JSMTE..10..035B} and is therefore inaccessible in leading order perturbation theory.
Unfortunately, these oscillations tend to mask the characteristic exponential decay of the false vacuum and prevent us from extracting easily the bubble nucleation rate. For larger values of the magnetic field, the particle excitation above the false vacuum disappears from the spectrum, however the vacuum structure drastically changes and the false vacuum disappears, which can be seen from the absence of any time intervals in which the exponential decay is valid. Note that in the case of the Ising model considered in Section \ref{sec:isingdecay}, there is no particle excitation on top of the two vacua, and so the vacuum decay can be seen for a finite range of the magnetic field enabling the extraction of the bubble nucleation rate.

For the case of the subleading magnetic ($\sigma'$) perturbation, the situation is similar as for the leading magnetic perturbation, with the different false vacuum survives for much larger values of the magnetic field $h'$, as discussed in~\cite{2022PhLB..82837008L}, and the same is true for the particle excitation on top of the false vacuum. As a result, the large oscillations persist and prevent the determination of the bubble nucleation rate. We return to discussing the effect of particle excitations on top of the false vacuum in Subsection~\ref{subsec:sigmaprime}.

\subsection{Subleading magnetisation deformation perturbed by the energy density operator}
\label{subsec:sigmaprime}
\begin{figure}
    \centering
    \includegraphics{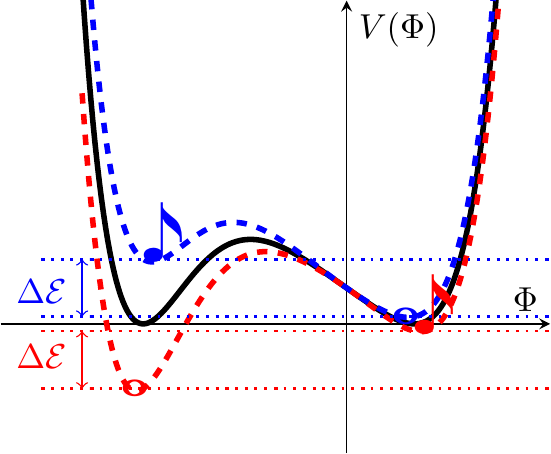}
    \caption{Qualitative Ginzburg--Landau potential in $\mathcal{A}_{\sigma'}$ (solid black) and its  $\epsilon$ deformations with $g>0$ (dashed blue) and $g<0$ (dashed red). The location of the false/true vacuum are denoted by {\large \eighthnote}/{\large \fullnote} respectively. The energy difference is also indicated between them.}
    \label{fig:A3epsGL}
\end{figure}
\begin{figure}
\begin{subfigure}{0.5\textwidth}
    \centering
    \includegraphics{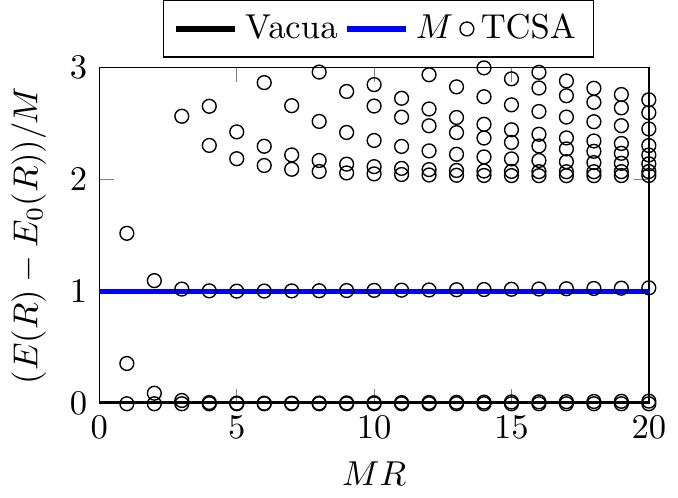}
    \caption{Finite volume spectrum of $H_{\sigma'}$. Note the presence of two degenerate vacua, however there is just a single neutral particle level since $\widetilde{K}_11$ is only present above one of them. The kink mass $M$ coincides with the exact mass gap \eqref{eq:a3gap} predicted by integrability (which is 1 in our units).}
    \label{fig:a3spect}
\end{subfigure}
\begin{subfigure}{0.5\textwidth}
    \centering
    \includegraphics{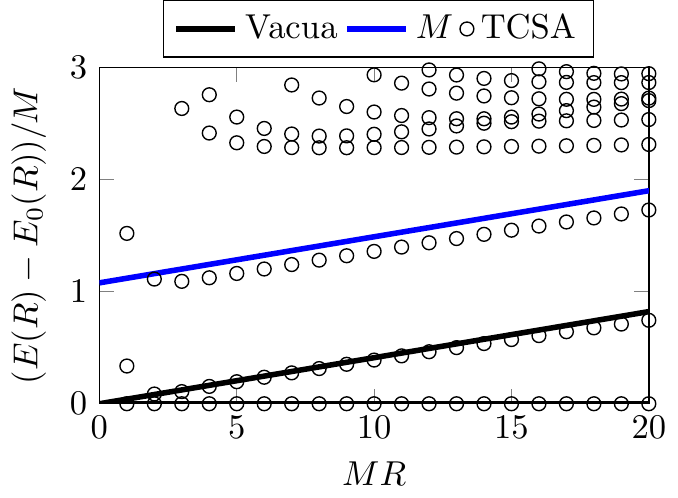}
    \caption{Finite volume spectrum of $H_{\sigma'}$ perturbed with $\epsilon$ with $g<0$, with the neutral excitation $\tilde{K}_{11}$ shown as a blue line running parallel to the false vacuum.}
    \label{fig:a3spectFerro}
\end{subfigure}
\begin{subfigure}{0.5\textwidth}
    \centering
    \includegraphics{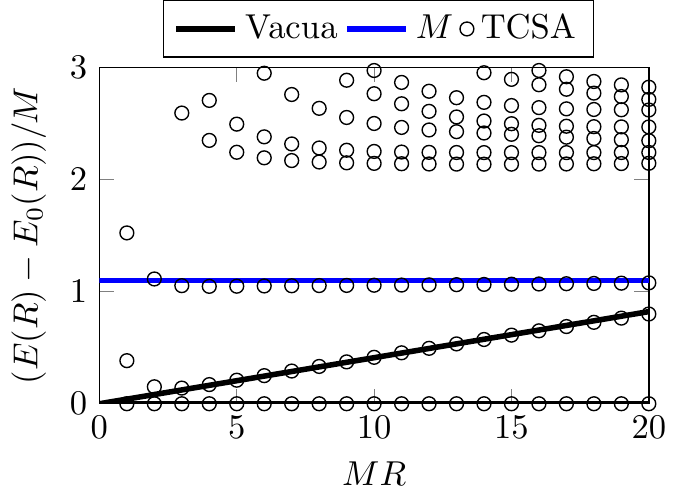}
    \caption{Finite volume spectrum of $H_{\sigma'}$ perturbed with $\epsilon$ with $g>0$, with the neutral excitation $\tilde{K}_{11}$ shown as a blue line running parallel to the true vacuum.}
    \label{fig:a3spectPara}
\end{subfigure}
\caption{Finite volume spectrum of the Hamiltonian $H_{\sigma'}$ (a) and its perturbation by $\epsilon$ for $g>0$ (b) and $g<0$ (c). The slope of the false vacuum shown by the corresponding black line is computed from~\eqref{eq:a3false}. For $g>0$ the neutral excitation is over the false vacuum and it therefore metastable, while for $g<0$ it is over the true vacuum and corresponds to a stable neutral particle excitation.}
\label{fig:a3spects}
\end{figure}
\begin{figure}
    \centering
    \includegraphics{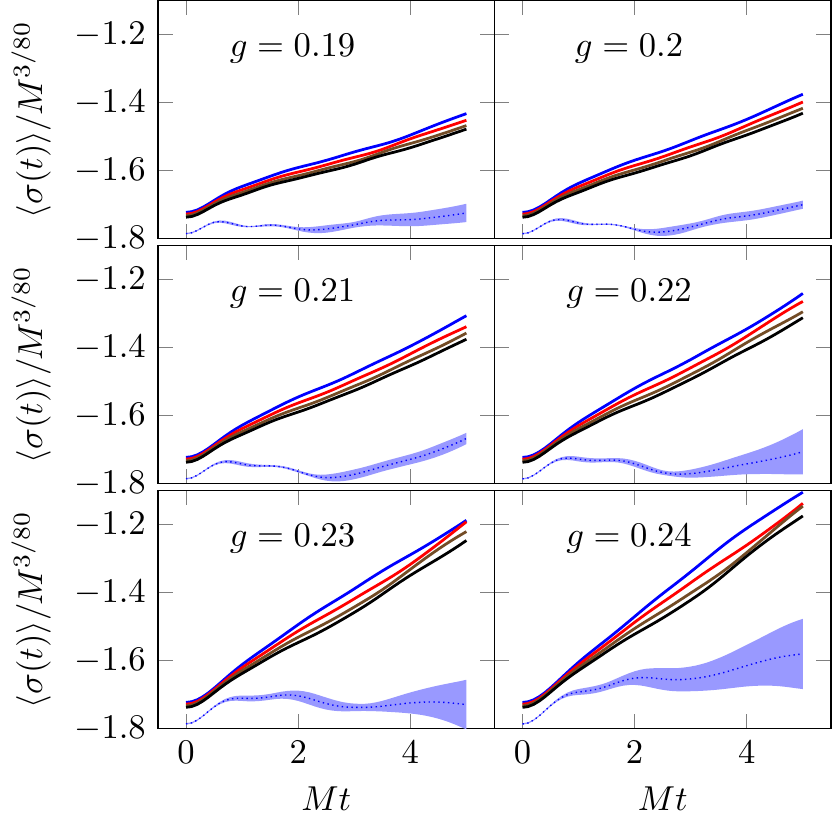}
    \caption{The time evolution of $\langle \sigma(t) \rangle$ in the $\mathcal{A}_{\sigma'}$ model perturbed by the energy density operator $\epsilon$, for various $g<0\rightarrow -g$ quenches from the ferromagnetic to the paramagnetic phase, ranging from $g=0.19$ to $0.24$ in dimensionless volume $MR=35$ and with time measured in units given by the kink mass $M$. Continuous lines are the raw TCSA data (blue, red, brown and black for cutoffs $15,16,17$ and $18$ respectively), while the dotted blue lines result from extrapolation in the cut-off, with shaded areas indicating the uncertainty of the extrapolation.}
    \label{fig:A3ParaSigmaVEVextrapol}
\end{figure}
\begin{figure}
    \centering
    \includegraphics{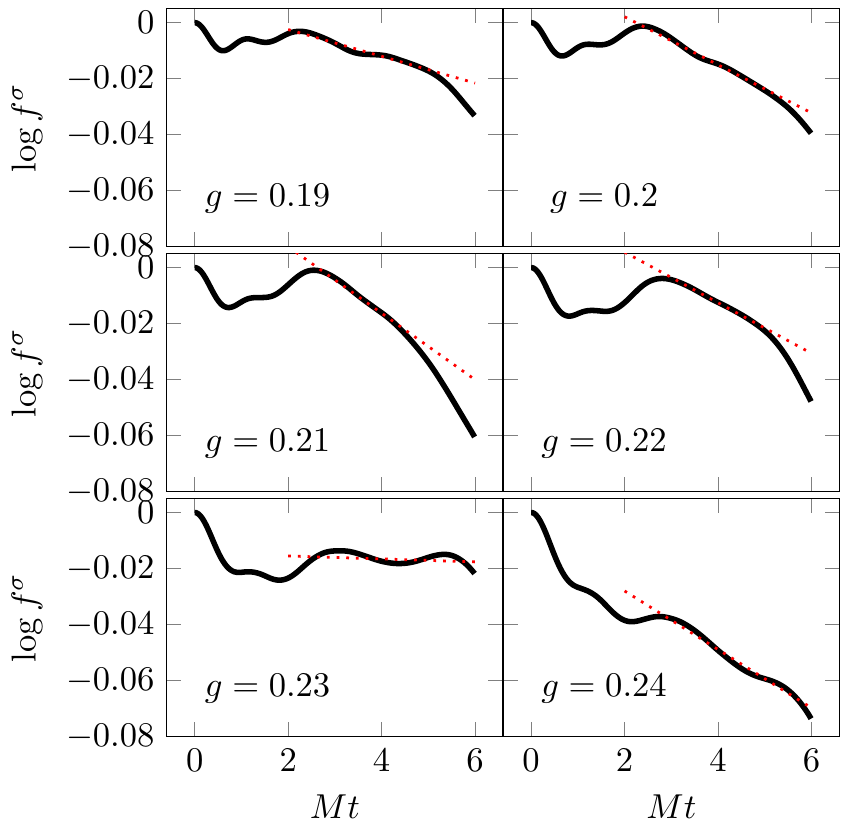}
    \caption{The time evolution of $\log f^{\sigma}$ (defined as in \eqref{eq:fsigma}) in the $\mathcal{A}_{\sigma'}$ model perturbed by the energy density operator $\epsilon$, for various $g<0\rightarrow -g$ quenches from the ferromagnetic to the paramagnetic phase, with $g$ ranging from $0.19$ to $0.24$, in dimensionless volume $MR=35$ and with time measured in units given by the kink mass $M$. We extract the bubble nucleation rate by fitting the slope of the linear part. In some cases (e.g. $g=0.23$) this is obviously not reliable, c.f. the discussion in the main text.}
    \label{fig:A3ParaSigmaVEV}
\end{figure}
\begin{figure}
    \centering
    \includegraphics{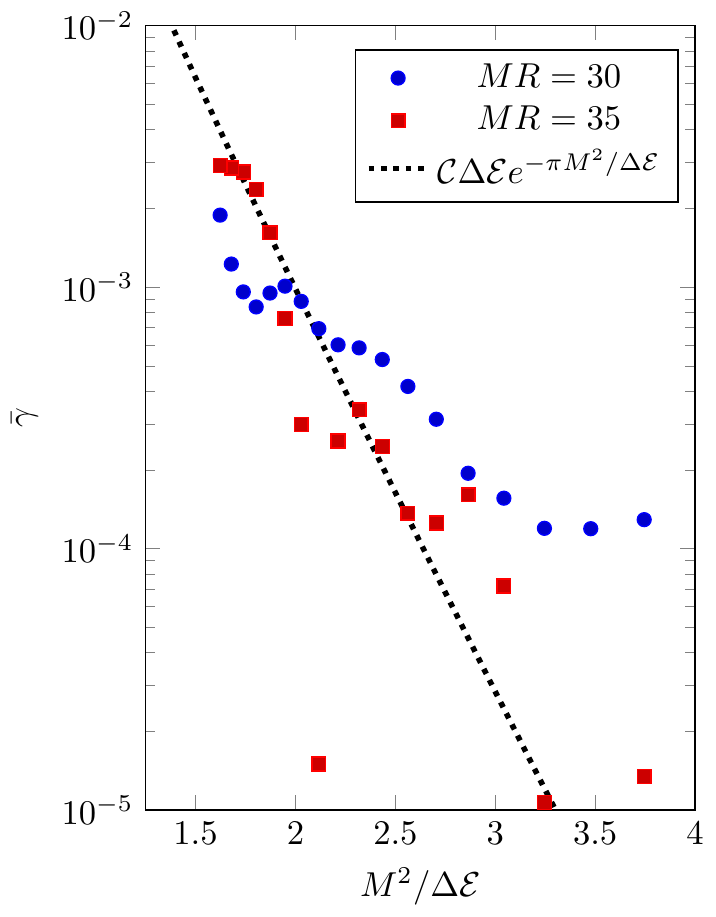}
    \caption{Dependence of the dimensionless bubble nucleation rate $\bar\gamma$ extracted from the time evolution of the expectation value of $\sigma$, on the latent heat $\Delta\mathcal{E}$ in different volumes in the $\mathcal{A}_{\sigma'}$ model perturbed by the energy density operator $\epsilon$, after quenching from the ferromagnetic phase to the paramagnetic phase. The value of the coefficient $\mathcal{C}$ was estimated as $44(7)$ for $MR=30$ and $87(18)$ $MR=35$; the theoretical curve shown in the plot uses the latter value. Due to the poor quality of the numerical determination of the nucleation rate, the difference between the two values carries no significance.}
    \label{fig:A3SigmaGamma}
\end{figure}

In the deformation of the tricritical Ising model with the subleading magnetisation ~\eqref{eqn:subleadingmagn}, the vacuum degeneracy can be lifted by adding the thermal operator:
\begin{equation}
    H=H_{\sigma'} + g \int dx\epsilon (x),
    \label{eq:A3eps}
\end{equation}
where $g<0$ corresponds to the ferromagnetic, while $g>0$ to the paramagnetic phase. In both cases the latent heat is
\begin{equation}
\Delta \mathcal{E} = |g(\braket{\epsilon}_1-\braket{\epsilon}_0)|,
\label{eq:a3false}
\end{equation}
where~\cite{1998NuPhB.516..652F}
\begin{align}
\braket{\epsilon}_1&=2.0445\dots \left|h'\right|^{8/45}\,,\nonumber\\
\braket{\epsilon}_0&=-0.78093\dots \left|h'\right|^{8/45}\,.
\end{align}
The corresponding Ginzburg--Landau potential is illustrated in Fig.~\ref{fig:A3epsGL}. Once again, the time evolution is followed by evaluating the expectation value of the leading magnetisation operator $\sigma$. The relation of the kink mass $M$ to the coupling $h'$ is given in Eq. \eqref{eq:a3gap}.

As discussed in Subsection \ref{subsec:sigmapvacs}, the vacuum structure of the unperturbed theory has no global symmetry, which leads to a physical situation very different from that of thermal deformations of the Ising and tricritical models. Concerning vacuum decay, there are now two different scenarios depending on the sign of the perturbing coupling $g$. 

Somewhat surprisingly, despite the physical difference between the vacua, the theoretical result \eqref{eq:voloshin} predicts that, in the thin wall limit, the bubble nucleation rate is independent of the direction of the tunneling provided we compare the two directions for equal values of the latent heat $\Delta\mathcal{E}$. However, we know that such a  prediction must be corrected to \eqref{eq:Cvoloshin} which leaves in principle open the possibility that an asymmetry may arise from the coefficient $\mathcal{C}$. Nevertheless, in Appendix \ref{sec:asymmetric_ratio} we demonstrate that the two bubble nucleation ratio is independent in the thin wall limit when computed up to one-loop order using Coleman's instanton approach \cite{Coleman1977,Coleman1977-2}, and it is likely that this result persists to all orders.

Despite the fact that it is possible to argue that the thin-wall limit of the bubble nucleation rate is expected to be identical in the two directions, there is still a marked difference in the dynamics. This asymmetry can be seen from the dependence on the spectrum on the sign of the perturbing coupling $g$, illustrated in Fig. \ref{fig:a3spects}.
\begin{itemize}
    \item For $g<0$, there is a particle on the top of the false vacuum corresponding to $\tilde{K}_{11}$ depicted in \ref{fig:A3kink}, as shown by the corresponding spectrum in Fig. \ref{fig:a3spectFerro}. Similarly to the discussion in Subsection \ref{subsec:E7}, when quenching from $g>0$ to $-g$ this produces large oscillations which make impossible to extract the exponential decay of the false vacuum. 
    \item For $g>0$, there is no particle excitation over the false vacuum as shown on Fig. \ref{fig:a3spectPara}. Therefore the related oscillations are absent, and  the decay of the one-point function can be clearly identified for the quench from $g<0$ to $-g$. 
\end{itemize}

Even in the second case, the TCSA extrapolation procedure has errors comparable to or even larger then the cut-off dependence which it is supposed to be removed, as visible in Fig.~\ref{fig:A3ParaSigmaVEVextrapol}. Nevertheless, the extrapolated data still allow us for a crude estimate of the bubble nucleation rate, illustrated in Fig.~\ref{fig:A3ParaSigmaVEV}, with the results for $\gamma$ presented on Fig.~\ref{fig:A3SigmaGamma} for two different volumes. For the larger volume $MR=35$, they are qualitatively consistent with the theoretical expectation, apart from a few outliers corresponding to values where the exponential decay could not be fit reliably. For the smaller volume $MR=30$ the agreement is much less precise, so one cannot really draw any conclusions concerning the volume (in)dependence of the fitting coefficient $\mathcal{C}$.

\subsection{Vacancy density deformation perturbed by energy density}
\begin{figure}
    \centering
    \includegraphics{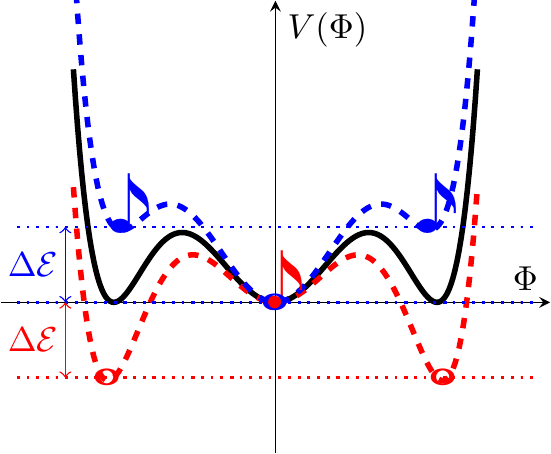}
    \caption{Qualitative Ginzburg--Landau potential in $\mathcal{A}_{\sigma'}$ (solid black) and its  $\epsilon$ deformations with $g>0$ (dashed blue) and $g<0$ (dashed red). The location of the false/true vacuum are denoted by {\large \eighthnote}/{\large \fullnote} respectively. The energy difference is also indicated between them.}
    \label{fig:A4epsGL}
\end{figure}

\begin{figure}
    \centering
    \includegraphics{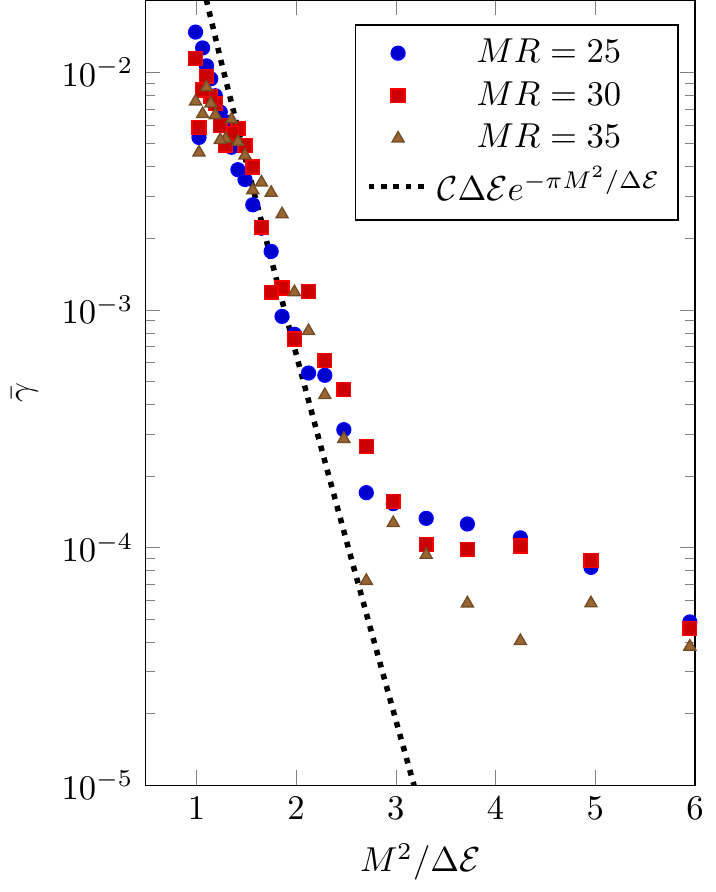}
    \caption{Dependence of the dimensionless bubble nucleation rate $\bar\gamma$ extracted from the time evolution of the expectation value of $\epsilon$, on the latent heat $\Delta\mathcal{E}$ in different volumes in the $\mathcal{A}_4$ model perturbed by $\epsilon$ after quenches from the phase with two stable vacua to the phase with a single stable vacuum, where the initial state is chosen to be the symmetric combination. The values obtained for the coefficient $\mathcal{C}$ were $0.8(2);0.6(1);0.7(2)$ in $MR=25;30;35$ respectively; the theoretical curve shown on the plot corresponds to the one with $\mathcal{C}$ obtained for $MR=35$. Note that quenches in the opposite direction i.e. starting from the phase with a single ground state lead to exactly the same results.}
    \label{fig:A4SigmaGammaSym}
\end{figure}

\begin{figure}
    \centering
    \includegraphics{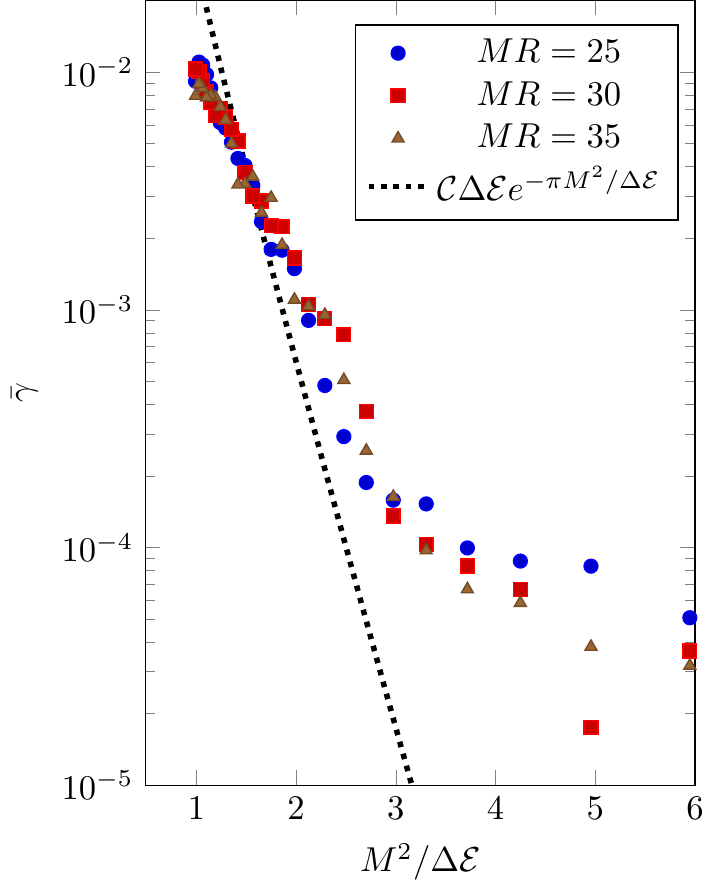}
    \caption{Dependence of the dimensionless bubble nucleation rate $\bar\gamma$ extracted from the time evolution of the expectation value of $\epsilon$, on the latent heat $\Delta\mathcal{E}$ in different volumes in the $\mathcal{A}_4$ model perturbed by $\epsilon$ after quenches from the phase with two stable vacua to the phase with a single stable vacuum, where the initial state is chosen to be the antisymmetric combination. The values obtained for the coefficient $\mathcal{C}$ were $0.6(1);0.7(2);0.6(1)$ in $MR=25;30;35$ respectively; the theoretical curve shown on the plot corresponds to the one with $\mathcal{C}$ obtained for $MR=35$.}
    \label{fig:A4SigmaGammaAsym}
\end{figure}

Here we consider the energy perturbation of the vacancy density deformation~\eqref{vacancypert}:
\begin{equation}
    H=H_{t}+g \int dx \, \epsilon(x).
    \label{eq:A4eps}
\end{equation}
The vacancy chemical potential $\mu$ in $H_{t}$ is taken to be negative. As discussed in Section \ref{sec:vacua}, the unperturbed model ($g=0$) has three degenerate ground state in the massive direction with kinks of mass $M$ which is given in terms of $\mu$ in Eq. \eqref{eq:A4kinkmass}. Compared to the simple $\mathbb{Z}_2$ vacuum structure of the thermal perturbation, the vacancy perturbation has a novel vacuum structure consisting of three degenerate vacua. The perturbation $\epsilon$ partially lifts the degeneracy due to the difference of the vacuum expectation value in the ground states labelled by $0,2$ and $1$~\cite{1998NuPhB.516..652F}:
\begin{align}
\braket{\epsilon}_{0,2}&=2.707495\dots \left|\mu\right|^{1/4}\,,\nonumber\\
\braket{\epsilon}_1&=-2.707495\dots \left|\mu\right|^{1/4}\,.
\label{eq:A4epsVEV}
\end{align}

As illustrated in Fig.~\ref{fig:A4epsGL}, switching on $g>0$ lifts up vacua $0$ and $2$, while $g<0$ lifts $1$, allowing the study of various decay scenarios. For the quench from $g<0$ to $-g$, the false vacua in finite volume correspond to even/odd combinations of $0$ and $2$, with a gap which vanishes exponentially with volume. Therefore, selecting the lower/higher of these finite volume levels as initial states corresponds to starting from the even/odd combination of the vacua $0$ and $2$, both of which then decay to the true vacuum corresponding to $1$. 

For the opposite direction of quenching from $g>0$ to $-g$ the initial false vacuum state corresponds to $1$ which then decays to the symmetric combination of the true vacuum states $0$ and $2$. It turns out that $g<0\rightarrow -g$ starting from the symmetric combination leads to the exact same time evolution as $g>0\rightarrow-g$ starting from $1$ due to the respective Hamiltonians being related by a $\mathbb{Z}_2$ symmetry; however, starting from the odd combination of $0$ and $2$ is a different scenario. As a result, it is sufficient to consider quenches $g<0\rightarrow -g$ to cover all physically different scenarios. We track the time evolution of the system by following the time evolution of the expectation value of the energy density operator $\epsilon$ and extract the bubble nucleation rate by the same method as for the Ising model discussed in Section \ref{sec:isingdecay}. Fortunately, for these quenches the cut-off extrapolation is reliable and, due to the absence of neutral particle excitations in the model $\mathcal{A}_t$, there are no oscillations to mask the exponential decay, allowing us for a precise determination of the bubble nucleation rate as long as the volume range satisfies the conditions discussed in Appendix \ref{subsec:TCSA_conditions}. As shown in Figs. \ref{fig:A4SigmaGammaSym} and \ref{fig:A4SigmaGammaAsym}, the value  numerically extracted from the time evolution matches very well the theoretical expectation \eqref{eq:Cvoloshin}, with the coefficient $\mathcal{C}$ displaying no significant dependence on the volume.

We close this section by observing that it is also possible to lift completely the degeneracy of the vacua perturbing with either of the leading/subleading magnetic fields, resulting in a novel interesting scenario of cascading decays. However, in all of these cases the cut-off dependence was found to be very strong, preventing the extraction of any reliable data regarding the false vacuum decay.

\section{Conclusions and outlook}\label{sec:conclusions}

In this work we have considered the decay of the false vacuum in 1+1-dimensional relativistic quantum field theories obtained as relevant perturbations of the scaling Ising and tricritical Ising conformal field theories. Formulating the vacuum decay as a quantum quench, we have followed the time evolution using the Truncated Conformal Space Approach (TCSA). This makes it natural to study the regime called the thin wall limit, since the magnitude of latent heat that can be handled is limited by the cut-off inherent in the TCSA method. In fact, this limitation is a bonus since vacuum decay allows us for an analytic treatment in the thin wall limit \cite{Coleman1977}, and indeed there are explicit predictions for the bubble nucleation rate for general 1+1-dimensional QFT by Voloshin~\cite{Voloshin1985}, as well as for the case of the Ising spin chain by Rutkevich~\cite{Rutkevich1999}. 

The first model considered in this work is the scaling Ising field theory. Here it is important to note that previous studies of vacuum decay \cite{2021PhRvB.104t1106L, 2022arXiv220515345S} performed using models in the same universality class have found that, while the dependence of the bubble nucleation rate on the latent heat was correctly predicted by the theoretical approaches, there is still a mismatch by a model dependent coefficient $\mathcal{C}$ defined in Eq. \eqref{eq:Cvoloshin}. In particular, for the $\varphi^4$ QFT studied in \cite{2022arXiv220515345S}, the quantity $\mathcal{C}$ was found to depend on the strength of the bosonic self-interaction. The Ising QFT studied here is eventually the straightforward scaling limit of the Ising chain considered by Lagnese et al.~\cite{2021PhRvB.104t1106L}, and our findings are fully consistent with those found on the spin chain, including the existence of the mismatch of the bubble nucleation rate by an overall constant coefficient. We also showed that the continuum limit of the Ising spin chain prediction \cite{Rutkevich1999} does not agree with the result claimed by Voloshin in the continuum limit \cite{Voloshin1985}. We also note that none of these predictions can match the numerically extracted coefficient of the bubble nucleation rate, therefore it is an interesting question to find a theoretical description which gives a fully correct prediction for the nucleation rate.

Then we have turned our attention to models obtained as deformations of the tricritical Ising field theory. These deformations provide  interesting situations since, apart from the simple scenario of vacua related to breaking of a $\mathbb{Z}_2$ symmetry as in the Ising class of models, there are also situations where the degenerate vacua are not related to any symmetry (this circumstance arises with the subleading magnetic deformation), or three degenerate vacua (associated to the vacancy density  deformation): altogether, these theories give rise to more general scenarios of vacuum decay.

In the case of the subleading magnetic deformation the interesting issue is the relation between the vacuum decay for the two possible directions. Voloshin's prediction~\cite{Voloshin1985} is that in the thin wall limit the nucleation rate is identical for the two directions, apart from a potential difference in the factor $\mathcal{C}$ which is necessary for matching it with the numerical results. As shown by an explicit calculation, the general instanton calculation shows that the two nucleation rates agree to one loop order, leading to the counter intuitive suggestion that they agree to all orders. However, limitations of the simulations prevented us from testing this idea with TCSA, due to the existence of a neutral excitation over one of the vacua which leads to large oscillations for one of the quench directions that obscures the decay dynamics.

For the case of the vacancy density deformation, lifting the deformation by the energy density operator leads to a vacuum structure with two of the three remaining degenerate, and so there are either two degenerate false and one true vacua or vice versa. It turns out that tunneling from the symmetric combination of false vacua $0$ and $2$ to true vacuum $1$ is identical to the opposite quench by flipping the sign of the energy coupling, while the tunneling from the antisymmetric combination of $0$ and $2$ to true vacuum $1$ is a different case. Nevertheless, the dependence on the latent heat is identical for these cases which is consistent with Voloshin's theoretical prediction~\cite{Voloshin1985}. The only potential difference between the rates of the symmetric/antisymmetric case is in the factor $\mathcal{C}$, however the numerics found no significant deviation between the extracted values $\mathcal{C}$. 

An interesting open direction for the future is to find out the reason for the mismatch between the theoretical predictions~\cite{Voloshin1985,Rutkevich1999} present in the literature.  Another interesting issue concerns  the improvement of the numerical simulations so that one can study the difference between tunneling directions in the case when there are asymmetric vacua, an effect which is certainly expected to be present beyond the thin wall limit. In addition, such improvements can help us realise scenarios with cascading false vacuum decays, which is the case for the vacancy deformation of the tricritical Ising models when vacuum degeneracy is lifted by the leading or subleading magnetisation operators.

\subsection*{Acknowledgments}
GM acknowledges the grant Prin $2017$-FISI. The work of ML was supported by the National Research Development and Innovation Office of Hungary under the postdoctoral grant PD-19 No. 132118. GT and ML were partially supported by the  National Research, Development and Innovation Office (NKFIH) through the OTKA Grant K 138606, and also within the Quantum Information National Laboratory of Hungary. This work was also partially supported by the CNR/MTA Italy-Hungary 2019-2021 Joint Project ``Strongly interacting systems in confined geometries”. 

\bibliography{vacuumdecay}
\bibliographystyle{utphys}
 
\appendix

\section{Vacuum decay in TCSA}
\subsection{General conditions for the TCSA simulation of vacuum decay}\label{subsec:TCSA_conditions}

To simulate the vacuum decay in TCSA, the following conditions need to be fulfilled:
\begin{enumerate}
\item The finite volume should be large enough to accommodate the critical bubble:
\begin{equation}
R\gg\frac{2M}{\Delta\mathcal{E}}
\label{eq:bubblesizelimit}\end{equation}
\item The energy cut-off $\Lambda$ should be high enough to allow for states with energies
comparable to the latent heat:
\begin{equation}
\Lambda\gg\Delta\mathcal{E} R
\label{eq:energylimit}\end{equation}
\end{enumerate}
For a given choice of the cut-off $\Lambda$ and the volume $R$, the above two conditions limit the range for the latent heat available for the simulation:
\begin{equation}
\frac{2}{MR}\ll\frac{\Delta\mathcal{E}}{M^{2}}\ll\frac{\Lambda}{M^{2}R}\,,
\end{equation}
which implies that ideally both the volume and the cut-off must be chosen as large as possible. However, the dimension of the Hilbert space grows as \cite{Cardy:1986ie}
\begin{equation}
    \propto \exp\left\{ 4\pi\sqrt{\frac{c\Lambda R}{24\pi}}\right\}\,,
\end{equation}
so a compromise must be struck that allows for a long enough range for the latent heat within the limitation of computer memory. 

Another important consideration is that due to the finite volume the simulation time is limited to $t<R$ to avoid revival effects, and in order to extract the nucleation rate with high enough precision it is necessary that the probability of decay during this time frame is not too small. 
\subsection{Cut-off dependence and extrapolation}\label{subsec:approximation}

The TCSA simulations were performed with the package developed in \cite{2022CoPhC.27708376H}, with the time evolution carried out by numerically solving the time-dependent Schrödinger equation using Matlab's \cite{MATLAB} ode45 solver. The dimensions of the truncated Hilbert spaces used in our calculations are listed in Table~\ref{tab:Hdims}.

\begin{table}
    \centering
    \begin{tabular}{|c|c|c|}
        \hline
        Truncation level & $\mathrm{dim}\mathcal{H}_{IM}$ & $\mathrm{dim}\mathcal{H}_{TIM}$ \\
        \hline
        \hline
        $13$ & $1994$ & $25040$ \\
        \hline
        $14$ & $3023$ & $41310$ \\
        \hline
        $15$ & $4476$ & $66628$ \\
        \hline
        $16$ & $6654$ & $106914$\\
        \hline
        $17$ & $9615$ & $168041$\\
        \hline
        $18$ & $14045$ & $263697$\\
        \hline
        $19$ & $20011$ & \\
        \hline
        $20$ & $28624$ & \\
        \hline
        $21$ & $40353$ & \\
        \hline
        $22$ & $56867$ & \\
        \hline
    \end{tabular}
    \caption{Hilbert space dimensions in the truncated spaces at different descendant level cut-offs.}
    \label{tab:Hdims}
\end{table}

The expectation values extracted from TCSA depend on the cut-off. For the case of expectation values in the vacuum or a low-energy eigenstate the leading cut-off dependence is of the form 
\begin{equation}
    \label{eq:extrapolation}
    \langle\mathcal{O}\rangle_\Lambda = \langle\mathcal{O}\rangle_\infty +  C\Lambda^{\nu_\mathcal{O}}
\end{equation}
where the exponent $\nu_\mathcal{O}$ is determined by the most singular term in the operator product expansion between the observable $\mathcal{O}$ and the interaction Hamiltonian density $V$
\begin{equation}
    \mathcal{O}(x)V(0)\sim \frac{A(0)}{|x|^{\alpha_{\mathcal{O}V}}}+\dots
\end{equation}
as $\nu_\mathcal{O}=\alpha_{\mathcal{O}V}-2$ \cite{2013JHEP...08..094S}. The $\nu_\mathcal{O}$ exponents for the different quantities with respect to various interaction terms used in the main text are summarized in Table~\ref{tab:exponents}.

\begin{table}
    \centering
    \begin{tabular}{|l|c|c|c|c|}\hline
         Model &$\mathcal{O}$ & $V$ & $A$ & $\nu_{\mathcal{O}}$\\
         \hline
         \hline
         IM,$\epsilon$,$\sigma$ & $\sigma$ ($1/16$) & $\epsilon$ ($1/2$) & $\sigma$ ($1/16$) & $-1$ \\
         \hline
         TIM,$\epsilon$,$\sigma$ & $\sigma$ ($3/80$) & $\epsilon$ (1/10) & $\sigma$ ($3/80$) & $-9/5$ \\
         \hline
         TIM,$\epsilon$,$\sigma'$ & $\sigma$ ($3/80$) & $\sigma'$ ($7/16$) & $\epsilon$ ($1/10$) & $-5/4$ \\
         \hline
         TIM,$t$,$\sigma$, & $\sigma$ ($3/80$)& $t$ ($3/5$) & $\sigma$ ($3/80$)& $-4/5$ \\
         \hline
         TIM,$t$,$\sigma$, & $\epsilon$ ($1/10$)& $t$ ($3/5$) & $\epsilon$ ($1/10$) & $-4/5$ \\
         \hline
         TIM,$t$,$\epsilon$ & $\epsilon$ ($1/10$)& $t$ ($3/5$) & $\epsilon$ ($1/10$) & $-4/5$ \\
         \hline
    \end{tabular}
    \caption{Leading exponents $\nu_\mathcal{O}$ used in the cut-off extrapolation of the time evolving expectation value $\langle\mathcal{O}(t)\rangle$, with the operators $V$ and $A$ corresponding to the leading exponent shown together with their (chiral) conformal weights. The models are specified by giving the UV CFT Ising (IM) / tricritical Ising (TIM), the deformation leading to the degenerate vacuum structure, and the perturbation lifting the degeneracy.}
    \label{tab:exponents}
\end{table}

Note that the above result only accounts for the cut-off dependence resulting in the static case for states much below the cut-off, and does not take into account cut-off dependence resulting from the time evolution with the truncated Hamiltonian. The latter can be partially improved by using the running coupling determined from the TCSA renormalisation group \cite{2006hep.th...12203F,2008JSMTE..03..011F,2011arXiv1106.2448G,2015PhRvD..91h5011R}. For a Hamiltonian of the form 
\begin{equation}
H=\frac{2\pi}{R}\left(L_{0}+\bar{L}_{0}-\frac{c}{12}\right)+\sum_{a}\lambda_{a}\int_{0}^{R}dx\Phi_{a}(x)
\end{equation}
where the perturbing fields $\Phi_a$ have conformal weights $h_a=\bar{h}_a$, the leading order RG equations in terms of the dimensionless couplings
\begin{equation}
 \tilde{\lambda}_{a}=\frac{\lambda_{a}R^{2-2h_{a}}}{(2\pi)^{1-2h_{a}}}   
\end{equation}
take the form \cite{2015JHEP...09..146L}
\begin{align}
&\tilde{\lambda}_{c}(n)-\tilde{\lambda}_{c}(n-1)=\nonumber\\
&\frac{1}{2n-d_{0}(r)}\sum_{a,b}\tilde{\lambda}_{a}(n)\tilde{\lambda_{b}}(n)C_{ab}^{c}\frac{n^{2h_{abc}-2}}{\Gamma(h_{abc})^{2}}\left(1+O(1/n)\right)\nonumber\\
&h_{abc}=h_{a}+h_{b}-h_{c}\label{eq:RG_eqs} 
\end{align}
where the $C_{ab}^{c}$ are the CFT operator product expansion coefficients:
\begin{equation}
\Phi_{a}(z,\bar{z})\Phi_{b}(0,0)=\sum_{c}\frac{C_{ab}^{c}\Phi_{c}(0,0)}{z^{h_{a}+h_{b}-h_{c}}\bar{z}^{\bar{h}_{a}+\bar{h}_{b}-\bar{h}_{c}}}\,,
\end{equation}
and 
\begin{equation}
    d_{0}(r)=\frac{R}{2\pi}E_0(R)
\end{equation}
is the vacuum scaling function given in terms of the finite volume vacuum energy $E_0(R)$, which can be estimated by its TCSA value at the starting cutoff for the RG run. In general, this prescription also gives a running coupling for the
identity, which leads to an additive renormalisation universal for
all energy levels that can be omitted in our simulations.

Nevertheless, it must be noted that during the course of time evolution further deviations accumulate from the truncation of the Hilbert space, resulting from the omission of states over the cut-off which increase with time and are not taken into account by the above improvements. As a result, albeit the extrapolation procedure based on \eqref{eq:extrapolation} can be very efficient \cite{Rakovszky_2016}, it is at best a useful heuristics whose validity must always to be verified. The rule of thumb we used in our calculation was to accept the extrapolated result when the fit error resulting from the extrapolation was significantly smaller than the cut-off dependence it was meant to eliminate.

\section{Continuum limit of the decay width for the Ising quantum spin chain}\label{sec:contlimit}

The quantum Ising spin chain
\begin{align}
H=&-J\sum_{n=1}^N\left(\sigma_{n}^{x}\sigma_{n+1}^{x}+h_{z}\sigma_{n}^{z}+h_{x}\sigma_{n}^{x}\right)
\\
&\sigma^a_{N+1}\equiv\sigma^a_1\nonumber
\end{align}
in the ferromagnetic phase with the transverse field $h_z<1$ is a lattice system with two vacua for $h_{x}=0$ which become degenerate for the thermodynamic limit $N\rightarrow\infty$. Its excitations are domain walls (kinks) which are free fermions with the dispersion relation
\begin{equation}
\omega(k)=2J\sqrt{1+h_{z}^{2}-2h_{z}\cos k}
\end{equation}
with gap $M=2J\left|1-h_{z}\right|$. Switching on the longitudinal field $h_{x}$ the vacuum degeneracy is lifted, and the theoretical decay amplitude of the false vacuum was computed by Rutkevich \cite{Rutkevich1999} with the result
\begin{equation}
\Gamma=\frac{\pi}{9}JN\left|h_{x}\right|\mu g\left(h_{x}\right)\exp\left\{ -\frac{\left|f\left(\theta_{0}\right)\right|}{\left|h_{x}\right|\mu}\right\} 
\label{eq:TIMdecayrate}\end{equation}
where 
\begin{align}
f\left(\theta\right)&=\frac{2}{J}\int_{0}^{\theta}dk\omega(k)\,,\nonumber\\ g\left(h_{x}\right)&=\text{Im}\cot\left[\frac{f(\pi)-i\pi\alpha}{2\left|h_{x}\right|\mu}\right]\,,
\end{align}
with $\alpha$ describing phenomenologically the decay rate of one-domain states,  $\mu=\left(1-h_{z}^{2}\right)^{1/8}$ is the spontaneous
magnetisation on the chain, while $\theta_{0}=i\left|\log h_{z}\right|$ is the zero of the function $\epsilon(k)$ in the upper half plane. 

The continuum limit of various quantities on the spin chain can be computed exactly \cite{Rakovszky_2016}. Introducing a lattice spacing $a=1/2J$ and the physical momentum $p=k/a$, and using $h_{z}=1-aM$, the dispersion relation becomes
\begin{equation}
\omega(p)=\frac{1}{a}\sqrt{1+h_{z}^{2}-2h_{z}\cos pa}=\sqrt{M^{2}+p^{2}}+O\left(a^{3}\right)
\end{equation}
which is the correct result for relativistic kinks with mass $M$. The continuum order parameter field is defined by
\begin{align}
\sigma(na)=&\bar{s}J^{1/8}\sigma_{n}^{x}\nonumber\\
&\bar{s}=2^{1/12}e^{-1/8}A^{3/2}\,,
\end{align}
where $A=1.2824271291\dots$ is Glaisher's constant. The continuum magnetic field is related to the lattice longitudinal field $h_x$ as
\begin{equation}
h=\frac{2}{\bar{s}}J^{15/8}h_{x}\,.
\end{equation}
These relations allow us to recover \eqref{eq:Isingsigmavev}
\begin{equation}
\bar{\sigma}=\bar{s}M^{1/8}\,.
\end{equation}
Note that
\begin{equation}
h\sigma(na)=2J^{2}h_{x}\sigma_{n}^{x}=\frac{1}{a}Jh_{x}\sigma_{n}^{x}
\end{equation}
resulting in the correct identification 
\begin{equation}
\int dxh(x)\sigma(x)=a\sum_{n}h\sigma(na)=J\sum_{n}h_{x}\sigma_{n}^{x}\,.
\end{equation}
Turning to the continuum limit of the nucleation rate, using $h_{z}=1-aM$ results in  $\theta_{0}=iaM$ and therefore
\begin{align}
f\left(\theta_{0}\right) & =\frac{2}{J}\int_{0}^{\theta_{0}}dk\omega(k)=4\int_{0}^{\theta_{0}}dk\sqrt{1+h_{z}^{2}-2h_{z}\cos k}\nonumber\\
 & \approx 4\int_{0}^{iaM}dk\sqrt{a^{2}M^{2}+k^{2}}\nonumber\\
 &=4i\int_{0}^{aM}d\kappa\sqrt{a^{2}M^{2}-\kappa^{2}}
 =i\pi a^{2}M^{2}
\end{align}
We also have $J\left|h_{x}\right|\mu=a\left|h\right|\bar{\sigma}$
where $\bar{\sigma}$ is the expectation value of the continuum order
parameter. Furthermore, the vacuum energy density difference is $\Delta\mathcal{E}=2\left|h\right|\bar{\sigma}$
so
\begin{equation}
\left|h_{x}\right|\mu=\frac{1}{J}a\frac{\Delta\mathcal{E}}{2}=\Delta\mathcal{E} a^{2}
\end{equation}
As a result, the exponential factor in the nucleation rate \eqref{eq:TIMdecayrate} has the continuum limit
\begin{equation}
\exp\left\{ -\frac{\left|f\left(\theta_{0}\right)\right|}{\left|h_{x}\right|\mu}\right\} =\exp\left\{ -\frac{\pi M^{2}}{\Delta\mathcal{E}}\right\}\,. 
\end{equation}
Turning to the prefactor, the $h_{z}\rightarrow1$ limit gives 
\begin{equation}
f\left(\pi\right)=4\int_{0}^{\pi}dk\sqrt{2-2\cos k}=8\,,
\end{equation}
resulting in 
\begin{equation}
g\left(h_{x}\right)=\text{Im}\cot\left[\frac{8-i\pi\alpha}{2\Delta\mathcal{E} a^{2}}\right]\rightarrow1\quad\text{as}\quad a\rightarrow0\,.
\end{equation}
Using that the physical volume is $R=Na$ leads to $N\left|h_{x}\right|\mu=V\Delta\mathcal{E} a$, which results in the following expression for the bubble nucleation rate
\begin{equation}
    \gamma=\frac{\Gamma}{R} =\frac{\pi\Delta\mathcal{E}}{18}\exp\left\{ -\frac{\pi M^{2}}{\Delta\mathcal{E}}\right\}\,,
\end{equation} 
which is to be contrasted with Voloshin's result \eqref{eq:voloshin}:
\begin{equation}
\gamma=\frac{\Delta\mathcal{E}}{2\pi}\exp\left\{ -\frac{\pi M^{2}}{\Delta\mathcal{E}}\right\}\,. 
\end{equation}
Clearly, the two results are identical except for a constant overall factor that is independent of both the kink mass $M$ and the latent heat $\Delta\mathcal{E}$.

\section{Ratio of the tunneling rates in the two directions in the case of asymmetric vacua}\label{sec:asymmetric_ratio}

Here we briefly consider the case of asymmetric vacua illustrated in Fig. \ref{fig:A3epsGL}, and show that the vacuum decay is independent of the direction in the thin-wall approximation, in the one-loop approximation. We assume that the model is described by a Ginzburg-Landau action containing a scalar field $\phi(t,x)$:
\begin{align}
\mathcal{A}=&\int dt dx \bigg[
\frac{1}{2}\left(\partial_t\phi\right)^2
-\frac{1}{2}\left(\partial_x\phi\right)^2-V_0(\phi)\nonumber\\
&- \eta \Delta V(\phi)\bigg]    
\end{align}
where in the case $\eta=0$ the potential has two degenerate vacua $V_0(\phi_+)=V_0(\phi_-)$. This degeneracy is assumed to persist at the quantum level, and it is only lifted by switching on the perturbing potential $\Delta V$. The $\pm$ index of the vacua is chosen so that $\phi_\pm$ becomes the false vacuum for $\eta$ positive/negative, respectively \footnote{Switching on the coupling shifts $\phi_\pm$ from their original values, however, this can be neglected at leading order in $\eta$; it can also be eliminated by a careful choice of the perturbing potential $\Delta V$.}. Using the semi-classical formalism, the bubble nucleation rate in the one-loop approximation is given by \cite{Coleman1977-2}
\begin{align}
\frac{\Gamma_{\pm}}{V}=&\frac{S_{0}}{2\pi}e^{-S_{0}}\left|\frac{\det'\left[-\partial_\tau^2-\partial_x^2+U''\left(\phi_{1}\right)\right]}{\det\left[-\partial_\tau^2-\partial_x^2+U''\left(\phi_{\pm}\right)\right]}\right|^{-1/2}\nonumber\\
&e^{-S^{(1)}(\phi_{1})+S^{(1)}\left(\phi_{\pm}\right)}\,,
\end{align}
where $\tau=-it$ is Euclidean time, $U=V_0+\eta\Delta V$ is the full potential, and $\phi_1$ is the classical instanton configuration, which solves the Euclidean equation of motion 
\begin{equation}
  \partial_\tau^2\phi_1+\partial_x^2\phi_1=V'(\phi)
\end{equation}
and has finite Euclidean action 
\begin{align}
S_0=&\int d\tau dx
\bigg[
\frac{1}{2}\left(\partial_\tau\phi_1\right)^2
\frac{1}{2}\left(\partial_x\phi_1\right)^2
+V_0(\phi_1)\nonumber\\
&+ \eta \Delta V(\phi_1)\bigg]
\end{align}
In the thin wall limit of small $\eta$ $\phi_1$ is given by
\begin{equation}
    \phi_1(\tau,x)=\phi_K(\rho-R)\quad,\quad\rho=\sqrt{\tau^2+x^2}\,,
\end{equation}
where $\phi_K(x)$ is the static kink solution at $\eta=0$. Denoting its mass by $M$, the radius of the bubble can then be determined as $R=2 M/\Delta\mathcal{E}$ where $\Delta\mathcal{E}\propto\eta$ is the difference in the energy densities of the false and the true vacua a.k.a. the latent heat. The thin wall limit means that the radius $R$ is much larger than the characteristic spatial extension of the kink solution $\phi_K$, which can always be achieved for sufficiently small $\eta$. The classical Euclidean instanton action in the thin wall limit can be easily computed following \cite{Coleman1977} with the result
\begin{equation}
    S_0=\frac{\pi M^2}{\Delta\mathcal{E}}
\end{equation}
and is independent of the sign of $\eta$, i.e., of the direction of tunneling. Finally, the contribution $S^{(1)}$ is the Euclidean one-loop counter term action.  

The ratio of the two tunneling amplitudes is then given by 
\begin{equation}
\frac{\Gamma_{+}}{\Gamma_{-}}=\left|\frac{\det\left[-\partial_\tau^2-\partial_x^2+V_0''\left(\phi_{-}\right)\right]}{\det\left[-\partial_\tau^2-\partial_x^2+V_0''\left(\phi_{+}\right)\right]}\right|^{-1/2}e^{S^{(1)}(\phi_{+})-S^{(1)}\left(\phi_{-}\right)}\,.
\end{equation}
Note that when the two vacua are related by a $\mathbb{Z}_2$ symmetry this ratio is trivially $1$. For the general case we can reason as follows. 
Since the $\phi_{\pm}$ are constant configurations, the relevant counter terms are the ones for the effective potential
\begin{equation}
S^{(1)}\left(\phi_{\pm}\right)=-\Omega\,\Delta V^{(1)}\left(\phi_{\pm}\right)\,,
\end{equation}
where $\Omega$ denotes a finite space-time box introduced to regulate the computation. This leads to the expression
\begin{align}
\log\frac{\Gamma_{+}}{\Gamma_{-}}= &-\frac{1}{2}\text{Tr}\log\left[-\partial_\tau^2-\partial_x^2+V_0''\left(\phi_{+}\right)\right]-\Omega\,\Delta V^{(1)}\left(\phi_{+}\right)\nonumber\\
 & +\frac{1}{2}\text{Tr}\log\left[-\partial_\tau^2-\partial_x^2+V_0''\left(\phi_{-}\right)\right]+\Omega\,\Delta V^{(1)}\left(\phi_{-}\right)\,.
\end{align}
The combinations
\begin{align}
&\frac{1}{2\Omega}\text{Tr}\log\left[-\partial_\tau^2-\partial_x^2+V_0''\left(\phi_{\pm}\right)\right]+\Delta V^{(1)}\left(\phi_{\pm}\right)=\nonumber\\
&\frac{1}{2}\int\frac{d^{2}k}{\left(2\pi\right)^{2}}\log\left[k^{2}+V_0''\left(\phi_{\pm}\right)\right]+\Delta V^{(1)}\left(\phi_{\pm}\right)
\end{align}
are just the renormalised $1$-loop contributions to the effective
potential evaluated at the field values $\phi_\pm$, which cancel due to the assumed exact degeneracy of the two (generally asymmetric) vacua. Therefore at the one-loop order one has
\begin{equation}
\Gamma_{+}=\Gamma_{-}\,.
\end{equation}
According to Voloshin's result Eq. \eqref{eq:voloshin} this equality is expected to hold in the thin wall limit to all loop orders. 
\clearpage
\end{document}